\documentclass{ws-p10x7}
\begin{document}
\title{Review of Final LEP Results\\ \mbox{\protect\rm\footnotesize or}\\A Tribute to LEP}

\author{J. Drees}

\address{CERN/University Wuppertal \\E-mail: Jurgen.Drees@cern.ch}

\twocolumn[\maketitle\abstract{ After a comment on the performance
of LEP some highlights of the LEP1 and LEP2 physics programmes are
reviewed. The talk concentrates on the precision measurements at
the Z resonance, two fermion production above the Z, $W^+W^-$
production, ZZ production, indirect limits on the Higgs mass, LEP
contributions to the exploration of the CKM matrix, and on the LEP
measurements of $\alpha_s$.}]

\section{Introduction}

\subsection{A comment on the machine and the detectors}\label{subsec:machine}

LEP delivered the last beam on November $2^{nd}$ 2000. By now the
storage ring and the detectors are dismantled. What remains is the
LEP saga and a rich harvest of physics results. So far more than
1100 scientific papers have been published covering an enormous
range of physics. The main topics centre on the study of the
properties of the gauge and scalar bosons, on heavy fermions and
on searches for the Higgs boson and for new physics. Many analyses
are still continuing, 220 papers have been submitted to this
symposium by the LEP collaborations.

\begin{figure*}
\epsfxsize30pc \figurebox{18pc}{32pc}{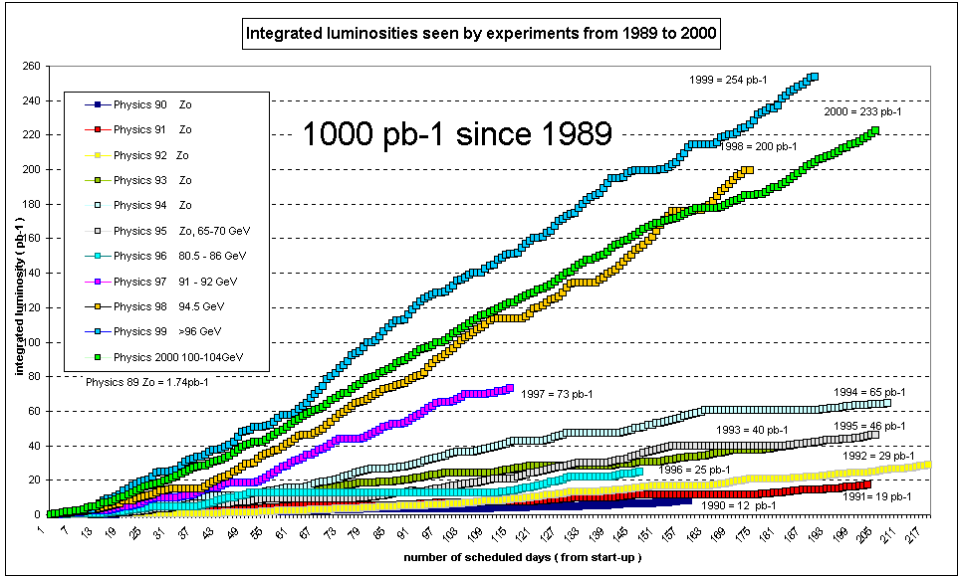}
\caption{Integrated luminosity delivered by LEP to each of the
four experiments from 1989 to 2000.} \label{fig:lumi}
\end{figure*}

The performance of LEP during the 12 years of operation can best
be illustrated by showing in Fig. \ref{fig:lumi} the integrated
luminosity as a function of time for each year. During the phase 1
where LEP operated in the vicinity of the Z resonance luminosities
up to 65 $pb^{-1}$ have been reached. After raising the energy the
luminosity increased to more than 200 $pb^{-1}$ per year. The
total luminosity delivered per experiment above $W^+W^-$
production threshold was about 700 $pb^{-1}$, while only 500
$pb^{-1}$ had been hoped for.

\begin{figure*}
\epsfxsize30pc \figurebox{16pc}{32pc}{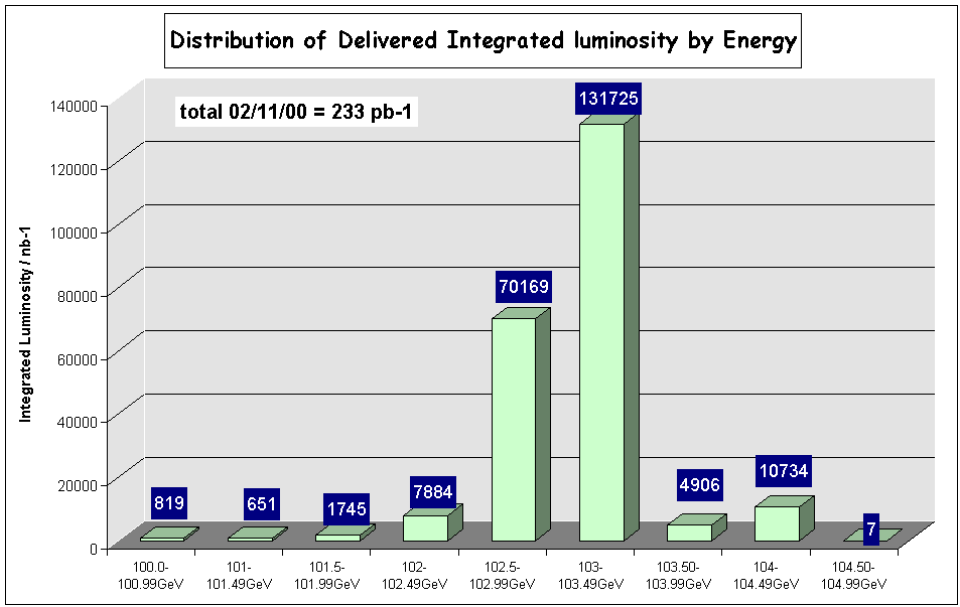}
\caption{Distribution of the integrated luminosity delivered to
each of the experiments in 2000 as function of the beam energy.}
\label{fig:energy}
\end{figure*}

In the hunt for the Higgs boson higher and higher energies were
achieved in 2000 which was a particularly good year for LEP. As
shown in Fig. \ref{fig:energy} a record beam energy of 104.4 $GeV$
was reached, much more than originally foreseen. In 200 days of
running more than 130 $pb^{-1}$ above 103 $GeV$ were delivered to
the experiments, 110 $pb^{-1}$ in the last 110 days mainly at beam
energies above 103 $GeV$.

Crucial for the success of LEP2 have been the superconducting
cavities. Let me quote here S. Myers\cite{myers}: {\it For
superconducting cavities the power needed is only proportional to
the $4^{th}$ power of energy. To operate LEP at 103 $GeV$ with
copper cavities (where the power would be proportional to
$E^8_{beam}$) would have needed 1280 cavities and 160 MW of power!
Impossible for many reasons.}

\begin{figure}
\epsfxsize120pt \figurebox{120pt}{120pt}{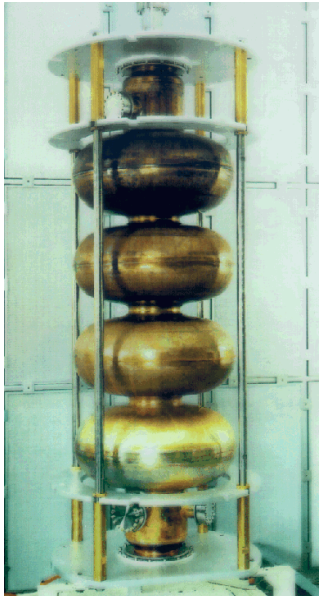}
\caption{A 4 cell Niobium coated cavity in the clean room.}
\label{fig:SC4}
\end{figure}

At the time when plans for superconducting cavities were developed
little was known about their performance\cite{pinkbook}. The final
success was due to a long term development programme, which
started already in 1980 together with outside laboratories,
pursuing the goal to reach thermal stability for 350 MHz niobium
coated copper cavities at reduced costs. In 2000 a total of 272 Nb
film and 16 Nb bulk cavities were installed. At 104 $GeV$ beam
energy an average accelerating field of 7.5 $MV/m$ at a quality
factor of Q $> 3 \times 10^{9}$ at 4.5 $K$ was achieved, much
better than the design value of 6 $MV/m$. More than 80\% of the
superconducting cavities had Q $\geq 2.5 \times 10^9$ even at 8
$MV/m$. Fig. \ref{fig:SC4} shows a 4 cell cavity with its typical
rounded structure.

A word on the four LEP detectors ALEPH, DELPHI, L3, OPAL. All
collaborations improved their detectors substantially during the
years of data taking, the most important improvements being:

1. The development of silicon micro vertex detectors for high
resolution secondary vertex measurements. The installation of
these detectors greatly improved the quality of heavy flavour
physics.

2. All experiments replaced their first luminosity detectors by
new high-precision detectors capable of measuring small angle
Bhabha scattering with an accuracy well below 0.1 \%.

The LEP Collaborations also created a new style of working
together, the LEP Working Groups, of which the Electroweak Working
Group (EWWG) is best known. These groups have the task to combine
the results obtained by the four LEP Collaborations and also by
the SLD Collaboration working at the SLAC $e^+e^-$ linear collider
SLC taking proper account of all systematic correlations between
the data.

\section{Precision at the Z}

\subsection{Determination of the Z Resonance Parameters}\label{subsec:zresonance}

\begin{figure}
\epsfxsize200pt \figurebox{200pt}{200pt}{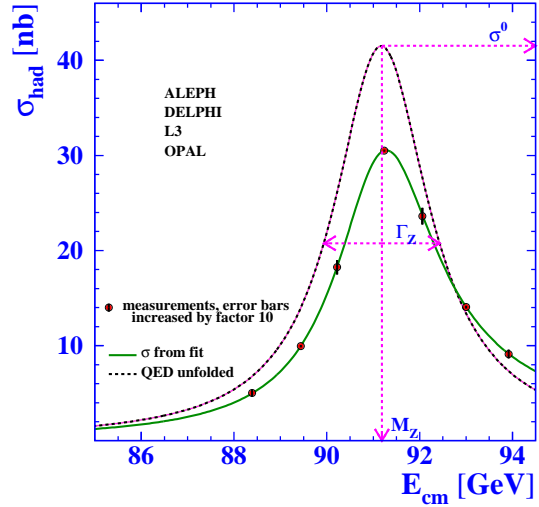}
\caption{The final hadronic cross-section as measured (solid line)
and QED deconvoluted (dotted line).} \label{fig:xsection}
\end{figure}

If one asks the question, what are the most important results from
LEP1, the answer has to be: the precision electroweak measurements
at the Z resonance. During the data taking periods from 1990 to
1995 the four experiments collected 15.5 million Z decays into
quarks plus 1.7 million decays to charged leptons corresponding to
an integrated luminosity of 200 $pb^{-1}$ per experiment. Fig.
\ref{fig:xsection} shows the hadronic cross-section measured by
the four collaborations as a function of the centre-of-mass
energy. Also shown is the cross-section after unfolding all
effects due to photon radiation. Radiative corrections are large
but very well known. At the peak the QED deconvoluted
cross-section is 36\% larger and the peak position is shifted by
-100 $MeV$. The figure illustrates the difference between the
measurements and the so-called pseudo-observables like $m_Z$,
$\Gamma_Z$, $\sigma^0_{had}$ which are averaged by the Electroweak
Working Group.

\begin{figure}
\epsfxsize190pt \figurebox{190pt}{190pt}{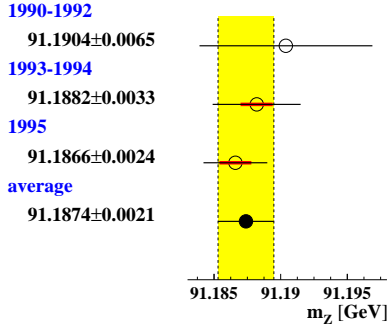}
\caption{$m_Z$ combined by EWWG for the different periods of data
taking.} \label{fig:consistency}
\end{figure}

The most impressive final result of the Z lineshape studies is the
$ 2 \times 10^{-5}$ accuracy for one of the most fundamental
constants of nature, the Z mass:
\begin{equation}
m_Z = {91.1874 \pm 0.0021 \; GeV}. \label{eq:mz}
\end{equation}
This precision cannot be exceeded by any one of the future
machines, not even with a GigaZ linear collider. Two essential
points have to be mentioned:\\ - The beam energy measurement using
the technique of resonant spin depolarisation plus careful control
of all machine parameters. Still the beam energy contributes 1.7
MeV to the total uncertainty of $ m_Z$. Fig. \ref{fig:consistency}
shows the consistency of the energy calibration for the different
data taking periods. \\ - The close cooperation with theory groups
essential for understanding radiative corrections with the
necessary accuracy.

The full set of nearly uncorrelated pseudo-observables used to
describe the precise electroweak measurements on the Z resonance
and combined by EWWG includes:

- The total Z width:
\begin{equation}
\Gamma_Z = {2.4952 \pm 0.0023}\; GeV. \label{eq:Gz}
\end{equation}

- The Z peak cross-section:
 \begin{equation}
\sigma^0_{had} \equiv
\frac{12\pi}{m_Z^2}\cdot\frac{\Gamma_{ee}\Gamma_{had}}{\Gamma_Z^2}.
\label{eq:sigmahad}
\end{equation}

- The ratios of the Z partial decay widths:
 \begin{equation}
 R^0_l \equiv \frac{\Gamma_{had}}{\Gamma_{ll}}\; with\, l=e, \mu, \tau. \label{eq:rlepton}
 \end{equation}

 \begin{equation}
 R^0_q \equiv \frac{\Gamma_{qq}}{\Gamma_{had}}\; with\, q=b, c, s. \label{eq:rquark}
 \end{equation}

- The pole forward-backward asymmetries:
  \begin{equation}
 A^{0,f}_{FB} \equiv \frac{3}{4}\mathcal A_e \mathcal A_f\; with\, \mathcal A_f \equiv
 \frac{2g_{Vf}g_{Af}}{g_{Vf}^2+g_{Af}^2}, \label{eq:afb0}
 \end{equation}
 for \(f = e, \mu, \tau, b, c, s\). Here \(g_{Vf}\) and \(g_{Af}\)
 denote the effective vector and axial-vector couplings to fermion {\it f}.

- The $\tau$ polarisation:
 \begin{equation}
 P_{\tau}(cos\theta)=- \frac{\mathcal A_{\tau}(1+cos^2\theta)+2\mathcal A_ecos\theta}
 {1+cos^2\theta+2\mathcal A_\tau \mathcal A_ecos\theta}.
        \label{eq:rf}
 \end{equation}

 Details on the final combination and an extended list of
 references can be found in\cite{zresonance}.
 The final measurements of Z line shape and of the leptonic forward-backward
 asymmetries performed by the
 four LEP Collaborations
 are documented in\cite{zALEPH,zDELPHI,zL3,zOPAL}. The measurements of
 the $\tau$ polarisation are obtained by the four
 collaborations by studying five $\tau$ decay modes\cite{tauALEPH,tauDELPHI,tauL3,tauOPAL}.

 Before summarizing the final results for the effective lepton
 couplings and the still preliminary results for the quark couplings
 I would like to mention two measurements of
 special interest. One of the questions asked by the LEPC before
 recommending approval of the experiments was: What is the
 expected accuracy for neutrino counting? I will come back to the
 answer given in 1982 at the end of the talk but here is the final
 measurement. The present best value results from the accurate
 measurement of \( \Gamma_{inv}/\Gamma_{ll} \) divided by
 \( \Gamma_{\nu\nu}/\Gamma_{ll} \), the latter evaluated from the
 Standard Model ( \( \Gamma_{inv}=\Gamma_Z - \Gamma_{had} - \Gamma_{ll}(3-\delta_\tau)
 \), $\delta_\tau$ corrects for the $\tau$ mass effect):
 \begin{equation}
 N_\nu = {2.9841 \pm 0.0083}. \label{eq:nnu}
 \end{equation}
 The value is consistent with 3 but 2 standard deviations below
 leaving room for a contribution of a new object to the invisible
 width of \( \Gamma_{inv}^x = -2.7^{+1.7}_{-1.5}\;MeV \).

 The second special quantity is the Veltman $\rho$-parameter.
 Assuming lepton universality $\rho$ can be
 determined from the measured leptonic width:
 \begin{equation}
 \rho^{lept}_{eff} = {1.0050 \pm 0.0010}. \label{eq:rho}
 \end{equation}
 The resulting $\rho^{lept}_{eff}$ value is found to be 5 standard
 deviations above the tree level of 1 thus proving the presence of
 genuine electroweak radiative corrections. It should be added
 that the experimental value agrees with the Standard Model
 expectation.

\subsection{Z couplings to charged leptons}\label{subsec:leptons}

\begin{figure}
\epsfxsize200pt \figurebox{200pt}{200pt}{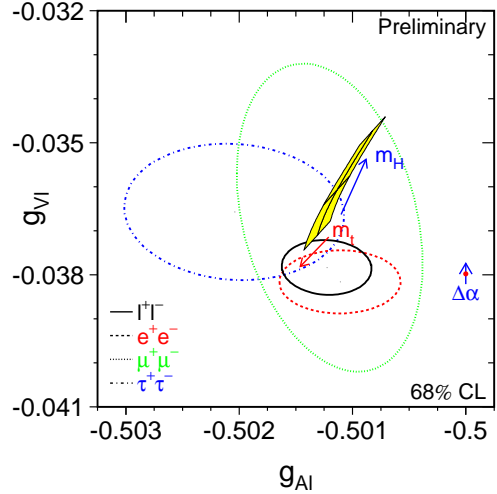}
\caption{The effective vector and axial-vector couplings for
leptons. The shaded area shows the prediction of the SM for
\(m_{top}=174.3\pm5.2\:GeV\) and \(m_H=300^{+700}_{-186}\:GeV\).
The arrows indicate increasing values of $m_{top}$ or $m_H$.}
\label{fig:leptoncouplings}
\end{figure}

By combining the measurements of the partial decay width of the Z
boson, which is proportional to the sum of the squares of the
vector and axial-vector couplings, with asymmetry measurements the
vector and axial-vector couplings can be determined separately.
For the three charged leptons the final results are presented in
Fig. \ref{fig:leptoncouplings}. It has to be noted that many data
enter this analysis. LEP contributes the measurements of the three
partial widths $\Gamma_{ll}$, the forward-backward asymmetries at
the Z (which yield \(\mathcal A_e,\mathcal A_{\mu},\mathcal
A_{\tau}\)), and the $\tau$ polarisation (\(\mathcal
A_{\tau},\mathcal A_e\)). SLD contributes the asymmetry for left
and right handed $e^-$ polarisation (yielding the most precise
individual measurement of \(\mathcal A_e\))\cite{alrSLD} and the
left-right forward-backward asymmetry for the three leptons
(\(\mathcal A_e,\mathcal A_{\mu},\mathcal
A_{\tau}\))\cite{afbpolSLD}. Assuming lepton universality the
result presented by the solid ellipse in Fig.
\ref{fig:leptoncouplings} is found. The comparison with the
Standard Model prediction shows the preference of the combined
lepton data for a low value of the Higgs mass.

\subsection{Z couplings to $b$ and $c$ quarks}\label{subsec:quarks}

\begin{figure}
\epsfxsize200pt \figurebox{160pt}{200pt}{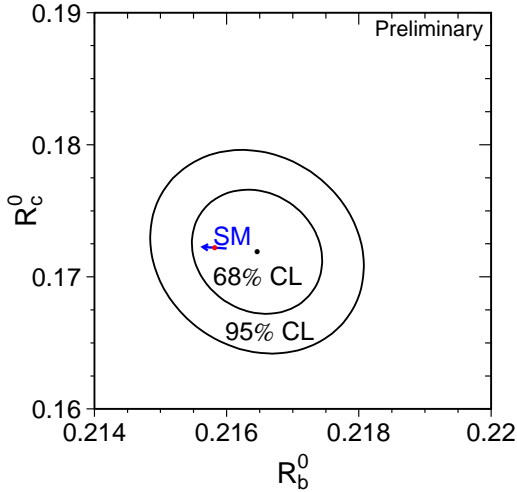}
\caption{Confidence level contours in the $R^0_c$, $R^0_b$ plane
obtained from the LEP and SLD data compared to the Standard Model
prediction for \(m_{top}=174.3\pm5.2\:GeV\).} \label{fig:rcrb}
\end{figure}

Information on the $b$ and $c$ quark couplings is obtained from
three types of observables: The ratios \(R^0_b\equiv
\Gamma_{b\bar{b}}/\Gamma_{had}\) and \(R^0_c\equiv
\Gamma_{c\bar{c}}/\Gamma_{had}\), which are measured by the LEP
Collaborations and by SLD, the forward-backward asymmetries
$A_{FB}^{0,b}$ and $A_{FB}^{0,c}$, which are measured at LEP, and
the direct measurements of \(\mathcal A_b\), \(\mathcal A_c\) by
SLD. Though the measurement of $R_b$ and $R_c$ is conceptually
simple, one has to separate an enriched sample of $b$ or $c$ quark
events from the bulk of the hadronic events, some problems have
been experienced in the past. A measurement of $R_b$, for
instance, requires extremely high quality of $b$ tagging, one has
to know the tagging efficiency and the background with sufficient
precision and one must control the correlations between the two
event hemispheres. The most precise measurements use double or
multi tag methods which allow the simultaneous experimental
determination of the tagging efficiency and the $b$ quark rate.
Combining the results of the five experiments results
in\cite{zresonance}:
\begin{eqnarray}
& R^0_b = {0.21646 \pm 0.00065}, \nonumber \\
& R^0_c = {0.1719 \pm 0.0031}. \label{eq:r0b} 
 \end{eqnarray} \\
The most recent $R_b$ work of the collaborations, all using a
lifetime tag based on micro vertex detector information plus
additional information from high $p_T$ leptons and the hadronic
structure of the event, can be found in references
\cite{rbALEPH,rbDELPHI,rbL3,rbOPAL,rbSLD}. An updated comparison
with the SM expectation is shown in Fig. \ref{fig:rcrb}. Obviously
the new $R_b$ and $R_c$ data agree with the prediction.

This is not the case for the $b$ forward-backward asymmetries. Two
new analyses of the pole asymmetry $A^{0,b}_{FB}$ by
ALEPH\cite{afbALEPH} and by DELPHI\cite{afbDELPHI} have been
submitted to this conference. These measurements are notoriously
difficult. One not only has to produce a high purity $b$ quark
sample, accurately control the background and understand the
hemisphere correlations, one also has to know whether a $b$ quark
or an anti-$b$ was produced in the forward hemisphere. Both
collaborations made optimal use of neural networks. ALEPH used a
neural network $b$-tag based on lifetime measurement, high $p_T$
leptons and event structure and obtains finally a $30\%$ increase
in the data sample. The $b$ hemisphere charge is estimated by an
optimal merging of the information from the primary and secondary
vertex charge, leading kaons and the jet charge. Their final
result is:

\begin{equation}
 A^{0,b}_{FB} = {0.1009 \pm 0.0031}. \label{eq:af0bALEPH}
 \end{equation}

\begin{figure}
\epsfxsize200pt \figurebox{160pt}{200pt}{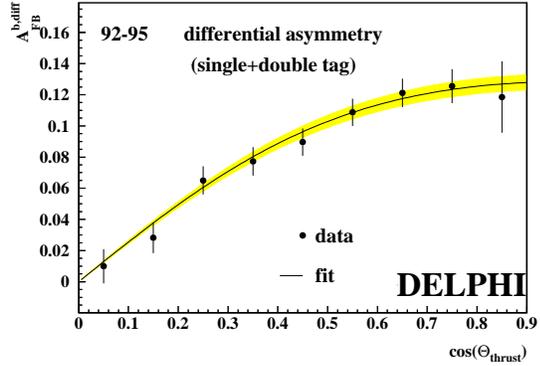}
\caption{The differential forward-backward $b$ asymmetry as
function of the polar thrust angle. The line is the result of a
fit with its statistical error indicated as a band.}
\label{fig:ZdiffafbDELPHI}
\end{figure}

DELPHI uses a very high purity $b$ sample ($96\%$), and a neural
network tag for the hemisphere charge combining the information
from vertex charge, jet charge, and from identified leptons and
kaons. Self calibration from double tagging is used to measure the
probabilities for $b$ or anti-$b$ tagging. The still preliminary
result is:
\begin{equation}
A^{0,b}_{FB} = {0.0997 \pm 0.0042}. \label{eq:af0bDELPHI}
\end{equation}
Fig. \ref{fig:ZdiffafbDELPHI} shows the differential $b$ quark
forward-backward asymmetry from the DELPHI single and double tag
data. The analysis includes all data collected from 1992 to 1995.

\begin{figure}
\epsfxsize210pt \figurebox{260pt}{210pt}{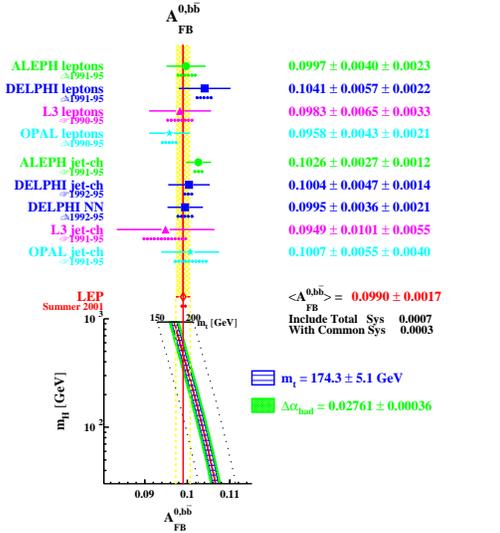}
\caption{$A^{0,b}_{FB}$ measurements from the LEP collaborations
using high $p_T$ leptons and various jet-charge techniques.}
\label{fig:Zallafb}
\end{figure}

These two measurements improve the accuracy of
$sin^2\theta^{lept}_{eff}$ evaluated from $A^{0,b}_{FB}$. However,
as in the past\cite{lp99SWARTZ}, there is still a significant
deviation of $3.3\, \sigma$ from the $sin^2\theta_{eff}$ value
determined from the lepton asymmetries. One then has to ask two
questions:

1. Are all LEP measurements consistent? This is clearly the case
as demonstrated in Fig. \ref{fig:Zallafb}, where the pole
asymmetries as measured by all LEP collaborations using different
analysis methods are collected. It should be remarked that the
numerical $A^{0,b}_{FB}$ values quoted in Fig. \ref{fig:Zallafb}
correspond to the measurements at the Z peak only. They do not
include measurements above and below the Z peak whose results are
included in Eq. (\ref{eq:af0bALEPH}). Including the off peak data
the average LEP value is:
 \begin{equation}
 A^{0,b}_{FB} = {0.0990 \pm 0.0017}. \label{eq:af0bLEP}
 \end{equation}
The error is dominated by statistics, the statistical error alone
being $\pm 0.00156$. The dominant contribution to the systematic
uncertainty is due to internal effects uncorrelated between the
experiments, the correlated systematic uncertainty is only $\pm
0.00039$.

2. Is the LEP result on \(\mathcal A_{b} =
\frac{4A^{0,b}_{FB}}{3A_e}\) consistent with the direct
measurements of \(\mathcal A_b\) from the polarised $b$ quark
forward-backward asymmetry? The results are \(\mathcal A_{b}(LEP
\, only) = {0.891 \pm 0.022}\) (last year $0.890 \pm 0.024$) and
\(\mathcal A_{b}(SLD) = {0.921 \pm 0.020}\). Both agree within 1
standard deviation.

\begin{figure}
\epsfxsize190pt \figurebox{190pt}{190pt}{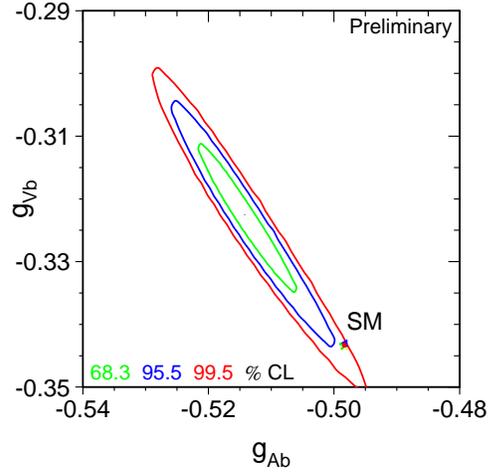}
\caption{LEP and SLD measurements of $g_{Vb}$ versus $g_{Ab}$
compared to the Standard Model prediction.}  \label{fig:gVbgAb}
\end{figure}

\begin{figure}
\epsfxsize190pt \figurebox{190pt}{190pt}{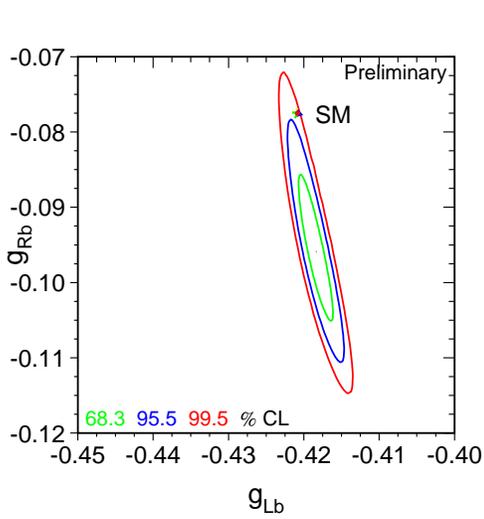}
\caption{LEP and SLD measurements of $g_{Rb}$ versus $g_{Lb}$
compared to the Standard Model prediction.}  \label{fig:gLbgRb}
\end{figure}

Using the information from all $b$ quark data, $R_b$,\,
$A^{0,b}_{FB}$, and \(\mathcal A_b\), one can separate the vector
and axial-vector couplings $g_{Vb}$,\, $g_{Ab}$ or the right and
left handed couplings $g_{Rb}$, $g_{Lb}$ respectively. They are
related by
\begin{eqnarray}
 g_{Rb} = (g_{Ab}-g_{Vb})/2, \nonumber \\  g_{Lb} = (g_{Ab}+g_{Vb})/2. \label{eq:gvalrb}
 \end{eqnarray} \\
The results are presented in Figures \ref{fig:gVbgAb} and
\ref{fig:gLbgRb}. Compared to the Standard Model prediction the
data show a deviation of about 3 standard deviations. The strong
anti-correlation in Fig. \ref{fig:gVbgAb} is due to the constraint
on the sum of the squares from the precise $R_b$ measurement. Fig.
\ref{fig:gLbgRb} shows that the deviation from the SM is mainly
for $g_{Rb}$.

\begin{figure}
\epsfxsize200pt \figurebox{200pt}{200pt}{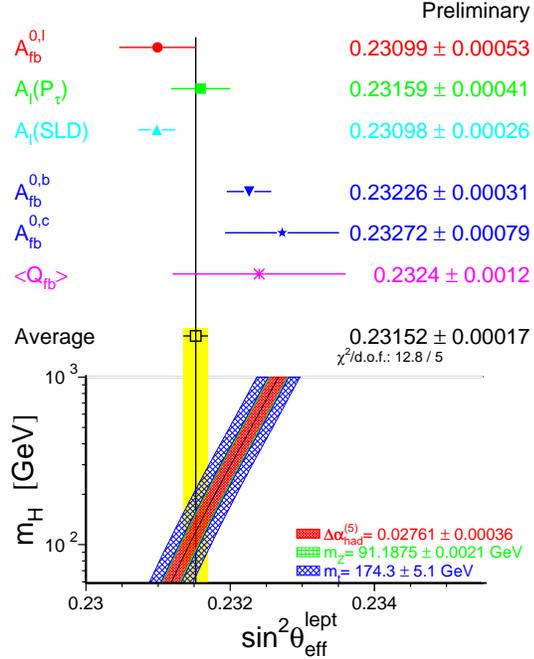}
\caption{The effective electroweak mixing angle
$sin^2\theta^{lept}_{eff}$ derived from data depending on lepton
couplings only (top) and from data depending on lepton and quark
couplings (bottom). Also shown is the prediction of the Standard
Model as a function of $m_H$. The band indicates the uncertainty
of the SM prediction due to the uncertainty of our knowledge on
$\Delta \alpha^{(5)}_{had}$, $m_Z$, and $m_t$. }
\label{fig:sin2theta}
\end{figure}

An update of the measurements of $sin^2\theta^{lept}_{eff}$ as
determined from lepton and quark data is presented in Fig.
\ref{fig:sin2theta}. While the lepton data prefer a small value of
$sin^2\theta^{lept}_{eff}$ and thereby a small Higgs mass the
quark asymmetries tend to larger $sin^2\theta^{lept}_{eff}$ and
$m_H$ values. Evaluating the average from the lepton data alone
yields:

\begin{equation}
sin^2\theta^{lept}_{eff}(leptons) = {0.23113 \pm 0.00021}.
\label{eq:s2thetaleptons}
\end{equation}
The corresponding average from the quark asymmetries is:
\begin{equation}
sin^2\theta^{lept}_{eff}(quarks) = {0.23230 \pm 0.00029}.
\label{eq:s2thetaquarks}
\end{equation}

The two values differ by 3.3 standard deviations. Presently this
deviation is unexplained. It could either be due to a statistical
fluctuation (the error of the most precise quark asymmetry
$A^{0,b}_{FB}$ is completely dominated by statistics), or due to
unknown sources of systematic errors (this is unlikely due to the
small systematic uncertainty correlated between the different
measurements of $A^{0,b}_{FB}$) or due to completely unexpected
new physics. However, one has to keep in mind that four of the
nine $A^{0,b}_{FB}$ measurements shown in Fig. \ref{fig:Zallafb}
are still preliminary. One should note that only the average of
lepton and quark $sin^2\theta^{lept}_{eff}$ measurements is
consistent with a Higgs mass of \(\mathcal O(100)\) $GeV$.

\section{Two Fermion Production above the Z}

\begin{figure}[t]
\epsfxsize200pt \figurebox{200pt}{200pt}{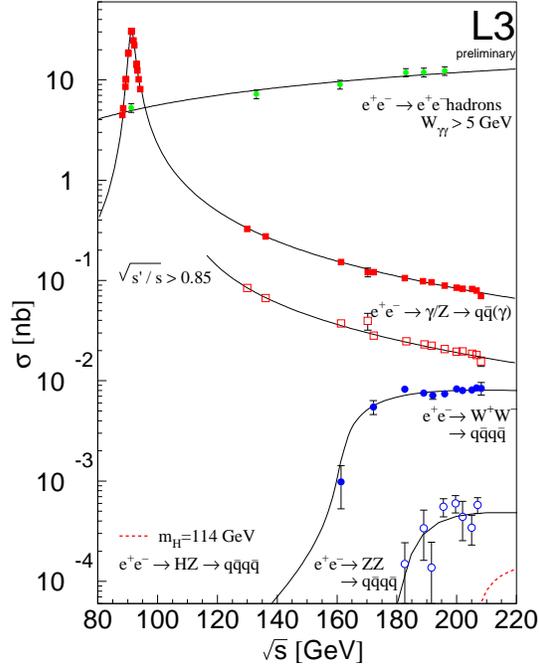}
\caption{Energy dependence of cross-sections in $e^+e^-$
annihilation. The data are from the L3 Collaboration. The
cross-sections for $e^+e^- \rightarrow q\bar q$ are shown for the
inclusive sample (full squares) and the non-radiative sample (open
squares). } \label{fig:l3xsections}
\end{figure}

\begin{figure}[t]
\epsfxsize190pt \figurebox{190pt}{190pt}{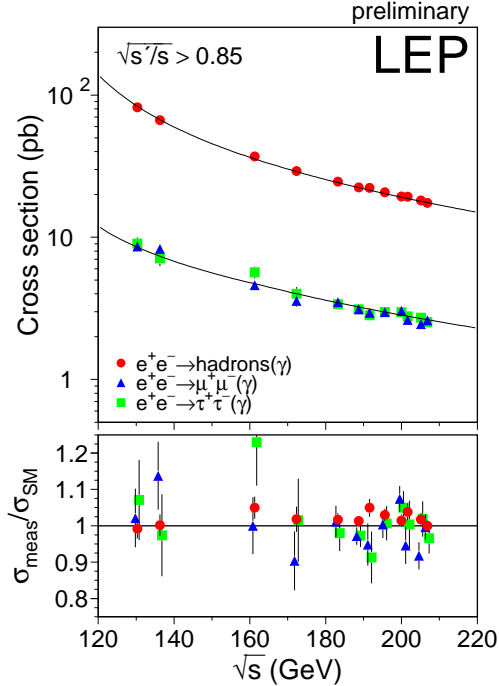}
\caption{Combined LEP measurements of the cross-sections for $q
\bar q$, $\mu^+ \mu^-$, $\tau^+ \tau^-$ production. The curves
show the SM expectation evaluated with ZFITTER. The lower part
shows the ratio data to SM prediction.} \label{fig:lepffxsections}
\end{figure}

\begin{figure}[t]
\epsfxsize190pt \figurebox{190pt}{190pt}{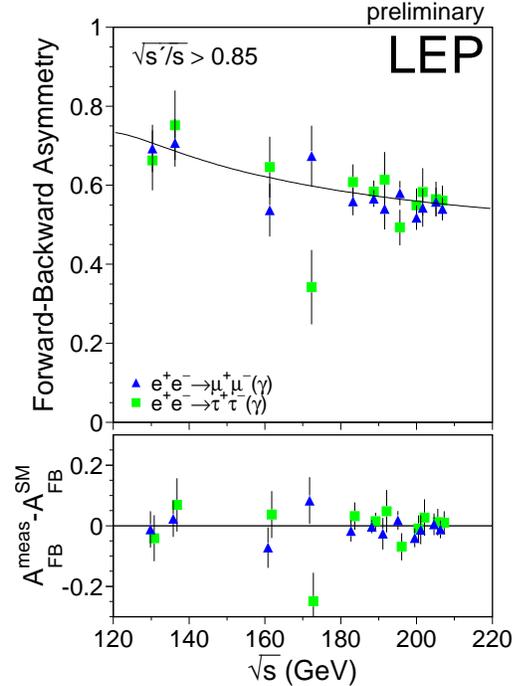}
\caption{Combined LEP results for the forward-backward asymmetries
for $\mu^+ \mu^-$ and $\tau^+ \tau^-$ final states. The curves
represent the SM expectation. The lower part shows the differences
between measurements and SM prediction.}
\label{fig:lepffasymmetries}
\end{figure}

Two fermion production at high energies provides a beautiful
laboratory for searching for new physics. Compared to other
processes the cross-section for $q \bar q$ production is still
high as shown in Fig. \ref{fig:l3xsections} prepared by the L3
Collaboration\cite{l3annual}, where the energy dependences of
cross-sections for various final states in $e^+e^-$ annihilation
are collected. At energies above the Z radiative processes are
important. Due to the large cross-section for radiative return to
the Z resonance only a fraction of the detected events have large
$s^{\prime}$, the square of the centre-of-mass energy transferred
to the $f \bar f$ final state. The Electroweak Working Group
defines the interesting non-radiative cross-section by
\(\sqrt{s^{\prime}/s} > 0.85 \) \cite{ffworkinggroup}. For this
cut the cross-sections for hadron, $\mu^+ \mu^-$, $\tau^+ \tau^-$,
$b \bar b$, $c \bar c$ production have been combined. Some results
are shown in Fig. \ref{fig:lepffxsections}. The lower part of the
figure presents the ratio of the data divided by the SM
prediction. Obviously the data are in agreement with the
prediction but one should notice that the hadronic cross-section
is $1.8\: \sigma$ high. The combined measurements of
forward-backward asymmetries for $\mu^+ \mu^-$ and $\tau^+ \tau^-$
final states are collected in Fig. \ref{fig:lepffasymmetries}.

The combined cross-sections and asymmetries and the results on $b$
and $c$ quark production have been used to study models with an
additional heavy neutral $Z^{\prime}$ boson. Limits for the
$Z^{\prime}$ mass have been obtained, for instance, for an E(6)
$\chi$ model \(m_{Z^{\prime}} > 0.68\) TeV or for the left-right
symmetric model \(m_{Z^{\prime}} > 0.80\) TeV. In both cases the
95\% confidence level lower limits are quoted and zero mixing with
the Z boson is assumed. It should be remarked that the LEP2 data
alone are not sufficient to constrain the mixing angle. But fits
including the LEP1 data of a single experiment are consistent with
zero mixing, see e.g. \cite{ffDELPHI}.

Many models for physics beyond the SM can be investigated in the
general framework of four-fermion contact interactions (analogous
to the low energy approximation of the weak force by Fermi
theory). Using the combined data, constraints have been placed on
the characteristic high energy scale $\Lambda$ describing the low
energy phenomenology of hypothetical new interactions. Limits for
contact interactions between leptons range from \(\sqrt{4
\pi}\Lambda/g
> 8.5\) to 26 TeV depending on the helicity coupling between
initial and final state fermions and on the sign of the
interference with the SM. Here $\it g$ is the coupling of the new
interaction. The corresponding limits for contact interactions
between leptons and $b$ quarks are \(\sqrt{4 \pi}\Lambda/g > 2.2\)
to 15 TeV, for leptons and $c$ quarks \(\sqrt{4 \pi}\Lambda/g >
1.4\) to 7.2 TeV.

Constraints have further been placed on the energy scale of
quantum gravity in compactified extra dimensions. Including data
from the Bhabha channel the typical result from the analysis of a
single experiment is \(M_s \geq 1\) TeV. Furthermore limits have
been set on the masses of leptoquarks. The $\gamma-Z$ interference
has been investigated in terms of the S-Matrix framework. In all
cases no deviations from the SM expectation have been observed.
Details on the two fermion analyses can be found
in\cite{ffALEPH,ffDELPHI,ffL3,ffOPAL,gravOPAL}. For a more
complete recent summary of the two fermion data and their
interpretation see\cite{holt}.

\section{$W^+W^-$ Production}

\begin{figure}
\epsfxsize190pt \figurebox{160pt}{190pt}{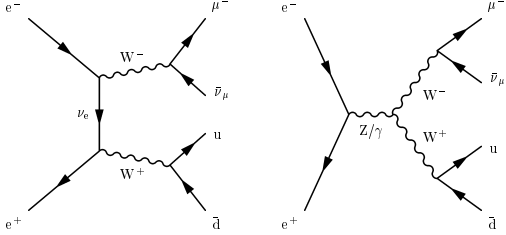}
\caption{CC03 diagrams for $W^+W^-$ production with subsequent
decay into $u\bar d$ and $\mu \bar \nu_\mu$.} \label{fig:CC03}
\end{figure}

Experimental studies of W-pair production have been a focus of the
LEP2 physics programme with two main goals: the measurements of
the W mass and the investigation of the structure of triple gauge
boson couplings. In $e^+e^-$ annihilation double resonant W pairs
are produced via the so-called CC03 diagrams shown in Fig.
\ref{fig:CC03}. Near threshold the cross-section is dominated by
the neutrino t-channel exchange. Contributions from the more
interesting s-channel exchange of a Z boson or a photon have been
measured at centre-of-mass energies from 172 to 209 $GeV$.

Each LEP experiment has finally collected about 10000 $W^+W^-$
events which are analysed in terms of five decay classes: fully
hadronic events where both W's decay into quarks, three
semileptonic decays and fully leptonic decays. In the SM the
branching ratio for the four quark class is 45.5\,\%, for each
semileptonic class 14.6\,\%, and for the fully leptonic class
10.6\,\%. Powerful tools to separate the four fermion events
originating from W production from the background have been
developed involving, for instance, neural networks. The efficiency
for WW selection is high, typically around 85\%, at very high
purity.

\begin{figure}
\epsfxsize200pt \figurebox{200pt}{200pt}{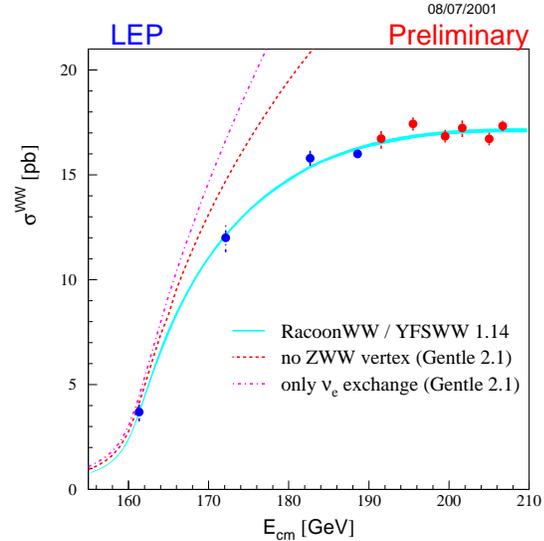}
\caption{The W-pair production cross-section as a function of the
centre-of-mass energy compared to the predictions of the Monte
Carlo generators RacoonWW and YFSWW.} \label{fig:wwxsection}
\end{figure}

The total CC03 cross-sections measured by the four collaborations
have been combined\cite{xsctLEPWW}, the results are summarised in
Fig. \ref{fig:wwxsection}. All experiments have published their
final results for centre-of-mass energies up to 189 $GeV$
\cite{fwwALEPH,fwwDELPHI,fwwL3,fwwOPAL}. The results for energies
up to 207 $GeV$ are still preliminary
\cite{prewwALEPH,prewwDELPHI,prewwL3,ffOPAL}. Inspection of Fig.
\ref{fig:wwxsection} immediately shows that all t- and s-channel
contributions are needed to understand the data. More subtle is
the comparison with predictions of the new four fermion generators
RacoonWW\cite{RacoonWW} and YFSWW\cite{YFSWW} with improved
radiative corrections. The calculations of both programmes are
based on the so-called double pole approximation for virtual $\cal
O(\alpha)$ corrections in resonant W-pair production plus all
other QED corrections needed for a 0.5\% accuracy. It is quite
remarkable that for $\sqrt{s} > 180$ $GeV$:

\begin{equation}
\sigma_{measured}/\sigma_{RacoonWW} = {1.000 \pm 0.009}.
\label{eq:sigmaratios}
\end{equation}

A very similar result is obtained for the calculation with YFSWW.

\subsection{Measurements of the W mass}\label{subsec:Wmass}

Even before crossing the W-pair threshold a precise value of the W
mass was evaluated from the LEP1 measurement of $m_Z$ using SM
relations. The updated indirect value obtained from a fit to all
data excluding the direct W mass measurements but including the
measured value of the top mass is \(m_W = 80.368 \pm 0.023\) $GeV$
\cite{ewwg}. The small error sets the scale for all direct
measurements. In the SM $m_W$ depends on electroweak loop
corrections. A recent complete two-loop calculation yields the
dependence on the top mass, the Higgs mass, and the QED induced
shift of the fine structure constant $\Delta \alpha$ as expressed
in Eq. (\ref{eq:mWtwoloop}).

\begin{table*}[t]
\begin{equation}
m_W = 80.3767 +
0.5235((\frac{m_t}{174.3})^2-1)-0.05613\,ln(\frac{m_H}{100})-1.081(\frac{\Delta
\alpha}{0.05924}-1) \pm ....  \label{eq:mWtwoloop}
\end{equation}
\end{table*}

In the Eq. (\ref{eq:mWtwoloop}) only the numerically most
important terms are shown, all masses are in $GeV$. For the
complete expression see\cite{Freitas}. An increase of $m_t$ will
increase, an increase of $m_H$ or $\Delta \alpha$ will decrease
the SM prediction for $m_W$. A significant deviation of a direct
measurement from the indirect value would indicate new physics and
the existence of new fundamental particles.

At LEP2 two independent and complementary methods have been used
to measure $m_W$. The first is based on the measurement of the
cross-section near threshold, which depends strongly on $m_W$.
Combining the measurements at a centre-of-mass energy of 161 $GeV$
the LEP groups obtain\cite{ewwg} \( m_W = 80.40 \pm 0.22\) $GeV$,
where the largest contribution to the total error is due to the
low event statistics. One should remark, however, that in
principle the threshold method can give a precise result, the
estimated error for a GigaZ Linear Collider\cite{lincol} is
\(\Delta m_W = 0.006\) $GeV$, supposing that radiative corrections
are controlled to this level.

\begin{figure}[t]
\epsfxsize160pt \figurebox{200pt}{200pt}{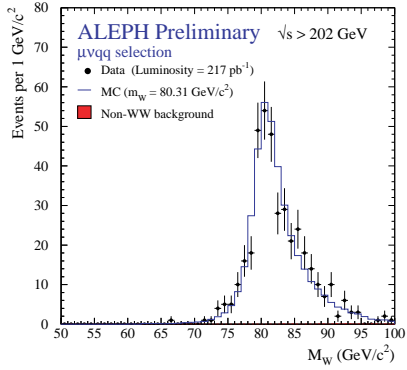}
\caption{Reconstructed invariant mass distribution from the ALEPH
experiment for the $q\bar q \mu \nu_{\mu}$ channel.}
\label{fig:qqmunuALEPH}
\end{figure}

\begin{figure}
\epsfxsize200pt \figurebox{150pt}{200pt}{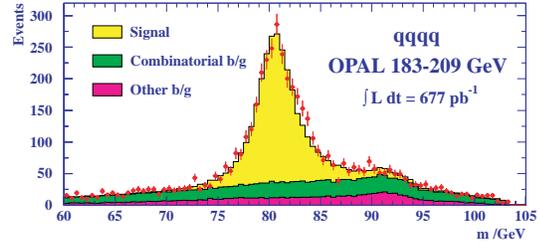}
\caption{Reconstructed W mass distribution for all OPAL \(W^+W^-
\rightarrow q\bar q q\bar q \) data from \(\sqrt{s} = 183\) to 209
GeV. The histogram shows the SM expectation for \(M_W = 80.42\)
GeV.} \label{fig:qqqqOPAL}
\end{figure}

At higher energies the W mass is directly reconstructed from the
invariant mass distribution of the decay products of the two W's.
Using constraints set by energy and momentum conservation clean
reconstructed mass distributions for the semileptonic and hadronic
decay channels are obtained. An example from the semileptonic data
taken at \( \sqrt{s}
> 202\) $GeV$ \cite{mwlargesALEPH} is reproduced in Fig.
\ref{fig:qqmunuALEPH}. Note that there is practically no
background in the $\mu \nu_{\mu} q\bar q$ channel. This also holds
for $e \nu_e q \bar q$ channel, the background in the $\tau
\nu_{\tau} q \bar q$ and 4q channels is small. The statistical
power of the data is illustrated in Fig. \ref{fig:qqqqOPAL}, where
the mass distribution for the fully hadronic channel as
reconstructed by the OPAL Collaboration is shown for all data
taken at $\sqrt{s}$ above 183 $GeV$ \cite{ffOPAL}. The data are
compared to the Monte Carlo prediction for $m_W = 80.42$ $GeV$.

From the measured masses in each event the final value of the W
mass is extracted by means of sophisticated analysis techniques,
which are somewhat different for the four experiments and in each
case require the comparison with a large number of Monte Carlo
events. ALEPH, L3, and OPAL use a reweighting technique to
determine the W mass, DELPHI uses a convolution technique. Details
on the analysis of the four experiments can be found in the final
publications for the data taken up to \(\sqrt{s} = 189\) $GeV$
\cite{mw189ALEPH,mw189DELPHI,mw189OPAL} or up to \(\sqrt{s} =
183\) $GeV$ \cite{mw183L3} and in more recent analyses contributed
to this conference\cite{mwlargesALEPH,mwlargesDELPHI,mwlargesL3}.

At present the precision of the combined result is limited by
systematic uncertainties. They are smallest for the mass values
extracted from semileptonic events. Here the total systematic
uncertainty is 29 $MeV$ with the largest contributions due to
fragmentation effects, beam energy uncertainty, detector
systematics, initial and final state photon radiation. The mass
determination from the fully hadronic events contains additional
uncertainties due to possible final state interactions between
quarks originating from the decay of different W's (colour
reconnection) or between hadrons (Bose-Einstein correlations).
Both effects may lead to distortions in the invariant mass
distribution, they are under study. Including such uncertainties
in a conservative way, a total systematic uncertainty of 54 $MeV$
is quoted for $m_W$ from fully hadronic events. The difference in
the masses obtained from the semileptonic and fully hadronic WW
decay channels is:

\begin{equation}
\Delta m_W(q \bar q q \bar q - q \bar q l \bar \nu) = {+ \,9 \pm
44 \: MeV}. \label{eq:mwdifference}
\end{equation}

\begin{figure}
\epsfxsize200pt \figurebox{200pt}{200pt}{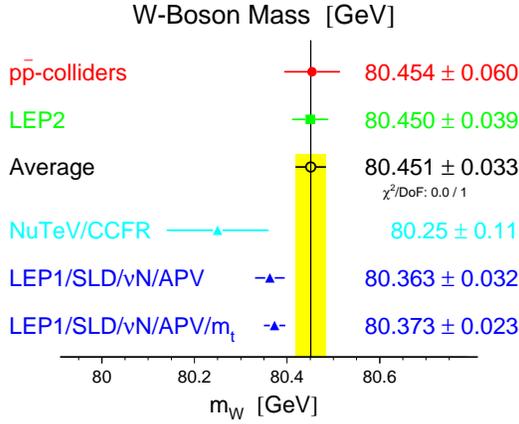}
\caption{Direct and indirect W mass measurements.}
\label{fig:mwWORLD}
\end{figure}

Combining all LEP measurements\cite{mwLEP} yields the nearly final
result:
\begin{equation}
m_W=80.450\pm0.026(stat.)\pm0.030(syst.)GeV. \label{eq:mwLEP}
\end{equation}
Here the weight of the fully hadronic channel in the combined fit
is only 26\%. All direct and indirect W mass measurements are
summarised in Fig. \ref{fig:mwWORLD}. Since not all LEP data are
included yet and studies of the final state interaction effects
continue it is hoped that the final LEP error will decrease to
about 35 $MeV$. There is still agreement between the indirect
determination from a fit including the measured top mass and the
direct measurements of $m_W$, but this year only within 1.9
$\sigma$.

The width of the W boson has also been measured at LEP: \(\Gamma_W
= {2.150 \pm 0.091 \: GeV}\). Within error there is good agreement
with the SM prediction.

\subsection{Charged Gauge Couplings}\label{subsec:CGC}

Measuring the specific form of the non-Abelian triple gauge boson
self-coupling $\gamma WW$ or $ZWW$ has been the second main goal
of W physics at LEP. Assuming electromagnetic gauge invariance,
charge conjugation and parity conservation and using also
constraints from low energy data reduces the number of couplings
from 14 in the most general case to three\cite{gounaris}: $g_1^Z,
\kappa_\gamma, \lambda_\gamma$ which have been most intensively
studied. Within the SM model these are given by 1,1,0 at tree
level. They are related to the magnetic dipole moment $\mu_W$ and
the electric quadrupole moment $q_W$ of the $W^+$:
\begin{eqnarray}
 & \mu_W & = \frac{e}{2m_W}(1+ \kappa_{\gamma}+\lambda_{\gamma}),
 \nonumber \\
   & q_W & = -\frac{2}{m_W^2}(\kappa_{\gamma}-\lambda_{\gamma}).
\label{eq:wmoments}
 \end{eqnarray}

A deviation of $\kappa_\gamma$ or $\lambda_\gamma$ from their SM
values would therefore prove the presence of anomalous
electromagnetic moments of the W boson and thus indicate
completely new physics in the boson sector. Results have been
derived using all available information from the total WW
production cross-section, the polar angular distribution of the
$W^-$, the $W^\pm$ helicities analysed via the fermion decay
angles, single W production \(e^+e^- \rightarrow e \nu W\), and
$\nu \bar \nu \gamma$ production. Within errors the measurements
agree with the SM expectation with the following precision
evaluated from one parameter fits to the combined
data\cite{cgcLEP}:

\begin{equation}
\delta g^Z_1 = \pm0.026,\: \delta \kappa_\gamma = \pm0.066,\:
\delta\lambda_\gamma = \pm 0.028. \label{eq:tgcprecision}
\end{equation}

Considering higher order effects the SM predicts small deviations
from the tree level values, e.g. $\Delta \kappa_\gamma \simeq
0.005$. Such small effects, however, are outside the scope of
present experimental verification.

In a more general approach the CP violating couplings have been
studied by ALEPH\cite{tgcgenALEPH} and OPAL\cite{tgcgenOPAL}.
Within errors no deviation from the SM has been observed. One
should mention that limits for the quartic charged gauge couplings
have been presented by ALEPH\cite{quarticALEPH},
L3\cite{quarticL3}, and OPAL\cite{quarticOPAL} albeit with large
errors. All results can be summarised by stating: no evidence has
been found for any anomalous W boson coupling.

\subsection{ZZ production}\label{subsec:ZZ}

\begin{figure}
\epsfxsize200pt \figurebox{200pt}{200pt}{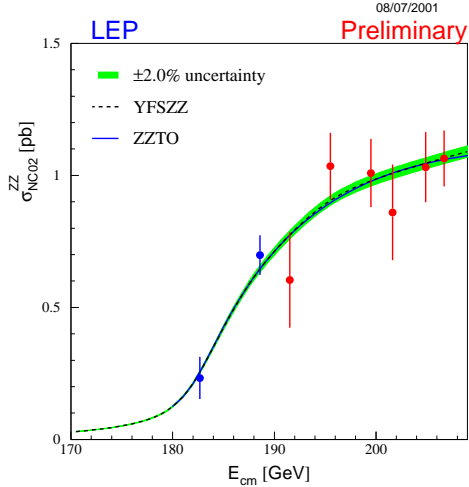}
\caption{LEP combined NC02 cross-sections. The curve shows the SM
expectation, the band corresponds to the $\pm 2\%$ uncertainty of
the prediction.} \label{fig:zzxsectionLEP}
\end{figure}

Measurements of ZZ production at $\sqrt{s} \geq 183$ $GeV$ allow
an investigation of a sector of the SM not tested before.
Deviations from the SM production cross-section, which is defined
by the NC02 diagrams involving only t- and u-channel electron
exchange, would be an indication for the existence of anomalous
neutral gauge couplings absent in the SM at tree level. The
ability to understand this process is also essential for the Higgs
boson search, where ZZ production forms an irreducible background.
All experiments have analysed ZZ decays into $q \bar q q \bar q$
(4 jets), $q \bar q \nu \bar\nu$ (2 jets plus missing energy), $q
\bar q l^+ l^-$ (2 jets plus 2 isolated leptons), and $l^+ l^- l^+
l^-$. New results have been submitted to this
conference\cite{zzxsectALEPH,zzxsectDELPHI,zzxsectL3,ffOPAL}.
Since the cross-section is only about 1 pb, a factor $\simeq 17$
smaller than the WW cross-section, the statistics is very limited.
The comparison of the energy dependence of the LEP combined data
to the SM prediction in Fig. \ref{fig:zzxsectionLEP}
\cite{xsctLEPWW} proves the agreement within the large errors of
the data.

The coupling of a virtual photon or Z boson to ZZ or Z$\gamma$
final states is not forbidden by fundamental principles. Non SM
contributions from the $\gamma^*ZZ$ or $Z^*ZZ$ vertex are
described by $f^{\gamma,Z}_i \; (i=4,5)$ couplings, from the
$\gamma^*Z \gamma$ or $Z^* Z \gamma$ vertex by $h^{\gamma,Z}_i \;
(i=1,4)$ couplings. Experimental tools to search for such
anomalous neutral triple gauge couplings are the measurement of
the total ZZ or $\gamma$Z cross-section (increase at high
energies?), the polar angle distribution of the produced Z or
$\gamma$ (deviations at large $\theta$?), and the $\gamma$ energy
distribution. New results submitted by all LEP
collaborations\cite{ngcALEPH,ngcDELPHI,ngcL3,ngcOPAL} have been
combined by the Electroweak Working Group\cite{cgcLEP}. For CP
conserving anomalous amplitudes a large interference with the SM
amplitude could arise. However, no evidence for anomalous neutral
couplings has been found. To give a few examples, the 95\%
confidence level limits for the CP conserving couplings
$f^\gamma_5$, $h^Z_3$ and $h^\gamma_3$ are:
\[ \begin{array}{l}
  f^Z_5 \:\:\:\: [-0.36,\: +0.39], \\  h^Z_3 \:\:\:\: [-0.20,\: +0.07],
\\  h^\gamma_3 \:\:\:\: [-0.049,\: +0.008]. 
\end{array} \]

\subsection{Consistency test of the SM}\label{subsec:consistency}

\begin{figure}
\epsfxsize200pt \figurebox{200pt}{200pt}{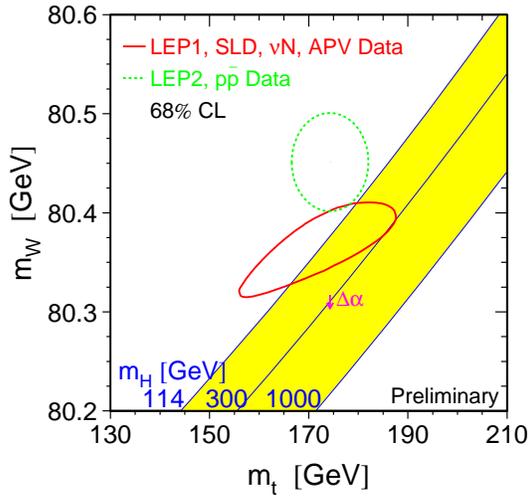}
\caption{Comparison of the indirect (full line) and the direct
(dotted line) measurements of $m_W$ and $m_t$. The diagonal band
shows the SM prediction for various values of the Higgs mass
ranging from 114 $GeV$ to 1000 $GeV$, $m_H \leq 114$ $GeV$ has
been excluded by direct searches.} \label{fig:mwmtcontours}
\end{figure}

\begin{figure}
\epsfxsize130pt \figurebox{200pt}{200pt}{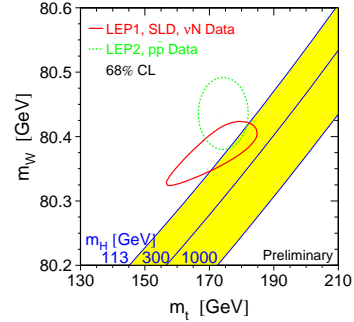}
\caption{Same as Fig. \ref{fig:mwmtcontours} but with the data
from summer 2000. The indirect result is obtained from an SM fit
to the LEP1, SLD, and neutrino nucleon data.}
\label{fig:contours2000}
\end{figure}

A consistency test of the SM can be performed by comparing the
indirect and the direct measurements of the W and the top quark
masses. In Fig. \ref{fig:mwmtcontours} the indirect contour has
been obtained from an SM fit to the data from LEP1, SLD, neutrino
nucleon scattering, and from atomic parity violation
experiments\cite{ewwg}. Both the direct and the indirect data
favour a low Higgs mass. The direct and the indirect measurements
still agree with each other though not as excellently as last year
(Fig. \ref{fig:contours2000}).

\begin{figure}
\epsfxsize200pt \figurebox{200pt}{200pt}{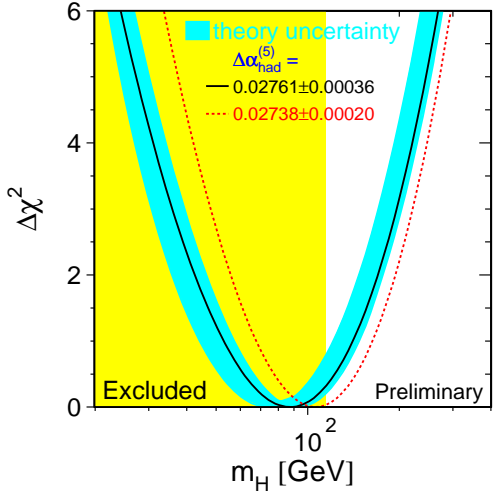}
\caption{$\Delta \chi^2 = \chi^2 - \chi^2_{min}$ as function of
the Higgs mass. The solid curve presents the result of the SM fit,
the band indicates the theoretical uncertainty. Also shown is the
Higgs mass $95\%$ CL exclusion limit from the direct search. The
dashed curve shows an SM fit assuming a lower value of $\Delta
\alpha^{(5)}_{had}(m_Z)$.} \label{fig:bluebands01}
\end{figure}

The experimental results of the direct searches for the Higgs
boson are discussed by G. Hanson\cite{hanson}, the theoretical
aspects by F. Zwirner\cite{zwirner}. With no significant Higgs
signal being observed, an indirect mass evaluation becomes again
important. Fig. \ref{fig:bluebands01} presents the updated version
of the traditional plot in form of a $\Delta \chi^2$ versus $m_H$
curve. The solid curve shows the result of the SM fit to all data
from LEP and SLD, the world data on $m_W$ and $m_t$, $sin^2
\theta_W$ from the neutrino experiments CCFR and NUTEV, the
measurements of atomic parity violation parameters, and also to
the new direct determination of $\Delta \alpha^{(5)}_{had}(m_Z)$
(the contribution of the 5 quarks to the running of the fine
structure constant $\alpha$) from\cite{burkhardt}. The fit
confirms the preference for a low Higgs mass. The 95\% confidence
level upper limit for $m_H$ is now 196 $GeV$. The dashed curve in
Fig. \ref{fig:bluebands01} is the result of a fit with $\Delta
\alpha^{(5)}_{had}$ from\cite{martin} but otherwise unchanged
input data and indicates the sensitivity of the $m_H$ prediction;
for details see\cite{ewwg}.

\begin{figure}
\epsfxsize180pt \figurebox{180pt}{180pt}{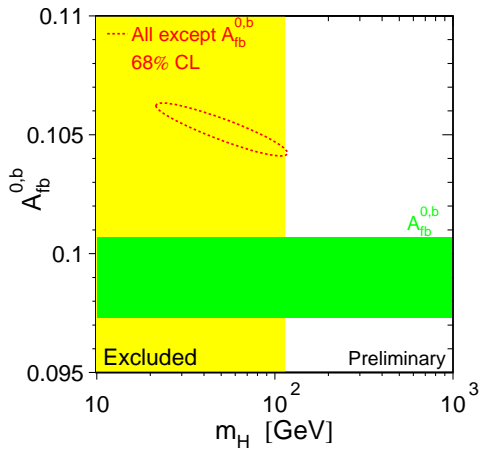} \caption{68\%
probability contour curve in the $(A^{0,b}_{FB},m_H)$ plane
obtained from an SM fit to all data except $A^{0,b}_{FB}$. The
direct measurement of $A^{0,b}_{FB}$ is shown as horizontal band
of width $\pm1$ $\sigma$. Also shown is the exclusion limit from
the direct Higgs search.} \label{fig:noafb}
\end{figure}

As discussed before the $b$ quark forward-backward asymmetry
deviates by about 3 $\sigma$ from its SM expectation. One may
therefore ask: what is the relative importance of including
$A^{0,b}_{FB}$ in the SM fit. The answer is given in Fig.
\ref{fig:noafb}, where the dotted contour line presents the 68\%
probability of the SM fit to all data except $A^{0,b}_{FB}$. The
preference for a low Higgs mass is even stronger, the one $\sigma$
contour is then completely excluded by the direct Higgs search.

\section{Contributions to the CKM Matrix}

LEP was part of the world wide effort to explore the structure of
the Cabibbo-Kobayashi-Maskawa quark-mixing matrix. From the
measurement of the W leptonic branching ratio one can determine $
V_{cs}$. More important, the determination of the CKM elements
$V_{ub}, V_{cb}$, and of the ratio $V_{td}/V_{ts}$ has been a
central part of the LEP B-physics programme. Strong points of the
LEP $b$ quark studies are: \\ - Large statistics, in total about 4
million Z $\rightarrow b \bar b$ decays, \\ - fast moving B
hadrons, the B hadron decay particles are well separated from the
QCD rest, \\ - tools for particle identification including
$K^{\pm}$, \\ - experience of 12 years of data analysis.

In the following only a few examples can be mentioned. A detailed
summary of combined B-physics results including the data from the
four LEP collaborations, from CDF and from SLD is
available\cite{bphyswg}.

\subsection{$| V_{cs}|$ from BR($W\rightarrow l \bar \nu$) }\label{subsec:vcs}

The leptonic branching fraction of the W boson is directly related
to the squares of the six CKM matrix elements not depending on the
t quark:

\begin{equation}
\frac{1}{3\, BR(W\rightarrow l\bar \nu)} = 1+[\,1+\frac
{\alpha_s(m_W)}{\pi}] \sum_{{i=u,c,}\above0pt{j=d,s,b}}
|V_{ij}|^2. \label{eq:wintolnu}
\end{equation} 

Taking the LEP average branching fraction as determined under the
assumption of lepton universality yields\cite{xsctLEPWW}:
\[ \sum |V_{ij}|^2 = 2.039 \pm 0.025  \] \\
consistent with the value of 2 expected from unitarity. With the
world average values for the other five CKM elements:

\begin{equation}
\mid V_{cs} \mid = 0.996 \pm 0.013. \label{eq:vcs}
\end{equation}

\subsection{Inclusive measurement of $|V_{ub}|$} \label{subsec:vub}

At LEP the measurement of $| V_{ub} |$ relies on the inclusive
reconstruction of the $b \rightarrow ul\bar \nu$ fraction:

\begin{equation}
|V_{ub}|^2 = \frac{BR(B\rightarrow X_ul\bar \nu)}{\gamma_b
\,\tau_b}, \label{eq:vubsquared}
\end{equation} \\
where $\tau_b$ is the average $b$ lifetime and $\gamma_b$ includes
QCD corrections and $b$ quark mass effects. Much progress has been
made during the last years as a consequence of both, improved
understanding of the theoretical uncertainties of $\gamma_b$ and
improved experimental analysis techniques\cite{bphyswg}.

\begin{figure}
\epsfxsize180pt \figurebox{180pt}{180pt}{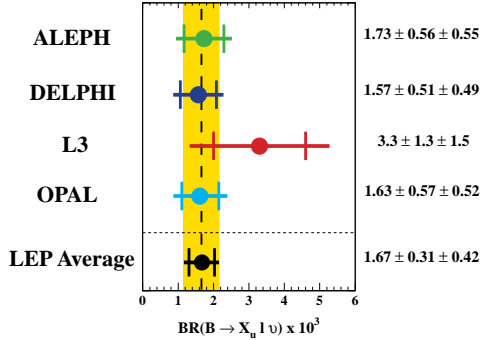}
\caption{Measurements of the branching ratio $B \rightarrow
X_ul\bar \nu $ by the four LEP experiments and the resulting
average. The first error is due to statistics and experimental
systematics uncorrelated between experiments, the second due to
all other systematic uncertainties.} \label{fig:vubLEP}
\end{figure}

Obviously it is very difficult to separate charmless $b$ decays
from the dominant $b \rightarrow c$ background. Several techniques
have been applied in earlier
publications\cite{vubALEPH,vubDELPHI,vubL3} based, for instance,
on inclusive analysis of semileptonic decays. In a new analysis
submitted to this conference the OPAL Collaboration uses 7
kinematic variables as neural net input in order to enrich the $B
\rightarrow X_ul\bar \nu$ sample\cite{vubOPAL}. All measurements
of the four collaborations are collected in Fig. \ref{fig:vubLEP}.

With the average branching ratio as determined by the LEP $V_{ub}$
Group:

\[BR(B \rightarrow X_ul^-\bar \nu_l) = (1.67 \pm 0.52)\times 10^{-3}\]
\\
and taking the world average B hadron lifetime \(\tau_b = (1.564
\pm 0.014)\)\,ps one finds:

\begin{equation}
|V_{ub}| = (4.04^{\, +0.59}_{\, -0.69})\times 10^{-3}.
\label{eq:vub}
\end{equation}

Here the error includes all theoretical uncertainties. The LEP
value of Eq. (\ref{eq:vub}) agrees very well with the most recent
measurement of the CLEO Collaboration\cite{cassel}. It should be
mentioned that the accuracy of $|V_{ub}|$ achieved at LEP is far
beyond of what was originally hoped for.

\subsection{$B^0_s - \bar B^0_s$ oscillations} \label{subsec:bsbsbar}

Much progress has also been made recently in the search for
$B^0_s$ oscillations. The main impact on the determination of the
CKM elements is explained in Eq. (\ref{eq:dmsdmd}):

\begin{equation}
\frac{\Delta m_s}{\Delta m_d} = \frac{m_{B_s}}{m_{B_d}}\; \xi^2\,
\frac{|V_{ts}|^2}{|V_{td}|^2}. \label{eq:dmsdmd}
\end{equation}

\begin{figure}[t]
\epsfxsize190pt \figurebox{190pt}{190pt}{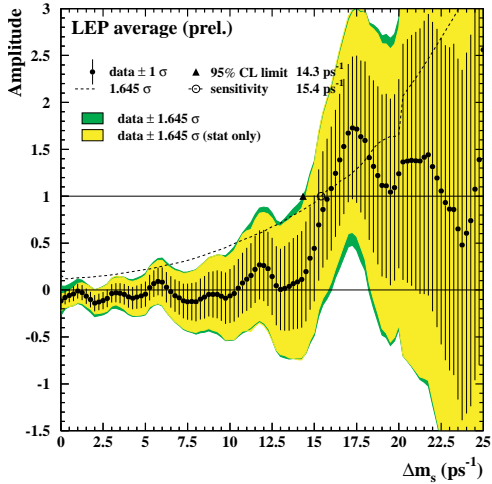}
\caption{Combined $B^0_s$ oscillation amplitude A as a function of
$\Delta m_s$. The $95\%$ CL limit derived from this spectrum is
marked by the small solid triangle.} \label{fig:dmsamplitude}
\end{figure}

In the ratio of the $B^0_s$ and $B^0_d$ mass differences $\Delta
m_s$ to $\Delta m_d$ many uncertainties cancel (see
e.g.\cite{pdg}) and the remaining non-perturbative quantity
$\xi^2$ is well known from lattice gauge theory:
\(\xi^2=1.16\pm0.05\) \cite{sachrajda}. No measurement of $\Delta
m_s$ has been performed yet, but upper limits have been set by
each experiment applying the so-called amplitude method. The idea
of the method is to replace the expression for the time dependent
probability that a produced $B^0_s$ is detected as $\bar B^0_s$ by

\begin{equation}
P(B^0_s \rightarrow \bar B^0_s) = \frac{1}{2}\,(1-{\bf A}\,
cos(\Delta m_s\,t))\,e^{-t/\tau_{B^0_s}} \label{eq:amplitude}
\end{equation}
\\
and then fit the amplitude A to the data for various fixed values
of $\Delta m_s$. Fig. \ref{fig:dmsamplitude} shows the amplitude
spectrum resulting from the combination of the spectra of all LEP
experiments\cite{bphyswg,boscgroup}. The combined spectrum
includes the new results from
DELPHI\cite{bsosc1DELPHI,bsosc2DELPHI} and from
OPAL\cite{bsoscOPAL}. From the LEP data in Fig.
\ref{fig:dmsamplitude} a $95\%$ confidence level lower limit of \(
\Delta m_s > 14.3 \: ps^{-1}\) is derived. Including the data from
SLD and CDF the present world limit increases to\cite{boscref}:
\begin{equation}
\Delta m_s > 14.6 \: ps^{-1} \: at \: 95\% \: CL.
\label{eq:dmsWorld}
\end{equation}
With the measured $B^0_s$ and $B^0_d$ masses and the world average
value of $\Delta m_d$ the limit for the ratio of the CKM elements
is now: \[|V_{td}|/|V_{ts}| < 0.22 \,.\]

\section{Contributions to QCD}

An important point to remember is that electroweak precision
quantities depend on the strong coupling $\alpha_s$. One of the
best known examples is the ratio of the Z partial decay widths
$R_{lept}$, which is known to \(\mathcal O(\alpha_s^3)\) as given
in Eq.(\ref{eq:rleptonQCD}). With the final value \( R^0_{lept} =
20.767 \pm 0.025\) (derived by assuming lepton universality) one
gets the result of Eq. (\ref{eq:alphas}). The advantage of
evaluating $\alpha_s$ from Eq. (\ref{eq:rleptonQCD}) is that
nonperturbative corrections are suppressed and the dependence on
the renormalization scale $\mu$ (which is often responsible for
the dominant uncertainty of $\alpha_s$ measurements) is small. All
theoretical uncertainties including the renormalization scale
uncertainty amount to only $+0.003, -0.001$, for details
see\cite{bethke}. Varying $m_t$ within $\pm 5\,$ $GeV$ and $m_H$
from 100 to 1000 $GeV$ leads to the additional small uncertainty
of $\pm 0.002$. A fit to all electroweak Z pole data from LEP and
SLD and to the direct measurements of $m_t$ and $m_W$ yields:
\(\alpha_s(m_Z) = 0.1183 \pm 0.0027\) \cite{ewwg}.

\begin{table*}[t]
\begin{equation}
R^0_{lept} =  \frac{\Gamma_{hadrons}}{\Gamma_{leptons}} =
19.934\,\{1+1.045 \frac{\alpha_s}{\pi} + 0.94
(\frac{\alpha_s}{\pi})^2 - 15(\frac{\alpha_s}{\pi})^3 \}
\label{eq:rleptonQCD}
\end{equation}
\end{table*}

\begin{table*}[t]
\begin{equation}
\alpha_s(m_Z) = 0.124 \pm 0.004(exp.) \pm 0.002(m_H,m_t)^{\:
+0.003}_{\: -0.001} (QCD). \label{eq:alphas}
\end{equation}
\end{table*}

\begin{figure}[t]
\epsfxsize167pt \figurebox{167pt}{167pt}{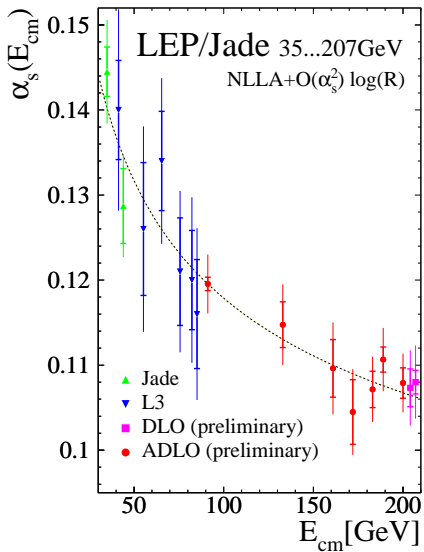}
\caption{Energy dependence of $\alpha_s$. The data are extracted
from the analysis of infrared safe hadronic event shape
distributions in the next-to-leading logarithmic approximation.
The dotted curve presents the expected running of $\alpha_s$.}
\label{fig:alljade}
\end{figure}

One may wonder whether these are the most reliable evaluations of
$\alpha_s(m_Z)$ using the LEP data. The problem is, however, that
the quoted results fully rely on the validity of the electroweak
sector of the SM. Small deviations can lead to large changes. It
is therefore necessary to measure $\alpha_s$ from infrared safe
hadronic event shape variables like jet rates, thrust, jet mass,
jet broadenings, etc. not depending on the electroweak theory.
Such studies have been performed by all LEP experiments, for more
recent publications
see\cite{alphasALEPH,alphasDELPHI,alphasL3,alphasOPAL}.
Measurements extracted by using resummed calculations in
next-to-leading logarithmic approximation (NLLA) matched to
\(\mathcal O(\alpha_s^2)\) calculations have been combined by the
LEP QCD Working Group\cite{qcdLEP}. As an example Fig.
\ref{fig:alljade} shows $\alpha_s$ values from fits to event shape
distributions at all LEP energies including measurements of the
JADE Collaboration at lower energies. A fit to the combined data
results in \(\alpha_s(m_Z) = 0.1195 \pm 0.0047 \), where the error
is almost entirely due to theoretical uncertainties
(renormalization scale). The figure also indicates to which extent
the running of $\alpha_s$ can be tested.

\begin{figure}[t]
\epsfxsize200pt \figurebox{200pt}{200pt}{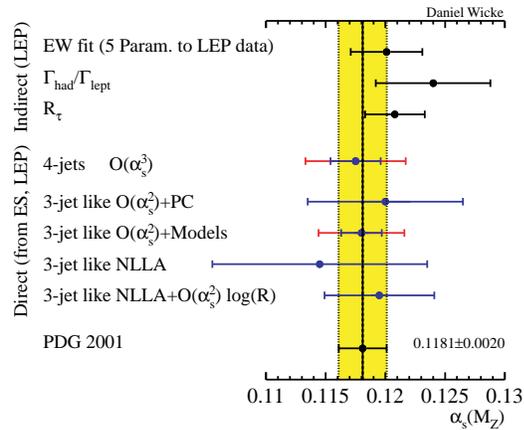}
\caption{Summary of $\alpha_s$ measurements at LEP compared to the
world average. The theoretical uncertainty for all 5 measurements
from event shapes (ES) is evaluated by changing the
renormalization scale $\mu$ by a factor of 2.}
\label{fig:allalphas}
\end{figure}

All LEP $\alpha_s$ measurements using a multitude of analysis
methods are collected in Fig. \ref{fig:allalphas} \cite{wicke}.
The three entries at the top present inclusive measurements for
which perturbative calculations are known in \(\mathcal
O(\alpha_s^3)\). One of the most precise measurements is obtained
from the ratio of the $\tau$ partial decay widths \(R_{\tau}=
\Gamma(\tau \rightarrow hadrons + \nu_{\tau})/\Gamma(\tau
\rightarrow e\bar \nu_e \nu_{\tau})\), the quoted value is
from\cite{pich}. The figure also includes the average values from
each of five different methods to extract $\alpha_s$ from hadronic
event shape distributions: four jet rates, 3 jet like observables
analysed in \(\mathcal O(\alpha_s^2)\) using either power
corrections or hadronic Monte Carlo generators for evaluating
hadronisation effects, three jet like observables analysed in pure
NLLA and in matched NLLA as mentioned above. All measurements
agree well with each other and with the world average.

Studying QCD at LEP has several advantages: the centre-of-mass
energy is high and well defined, jets are collimated, the
environment is clean, statistics is high enough to investigate
even rare topologies. In consequence more than 200 QCD papers have
been published till now including detailed investigations of
perturbation theory, hadronisation models, power corrections,
quark and gluon jet fragmentation, local parton-hadron duality,
soft gluon coherence etc. The experimental aspects are reviewed,
e.g. in\cite{biebel,bethkelepfest,duchesneau}. Of the many new QCD
studies contributed by the LEP Collaborations to this conference
only few can be briefly mentioned, for instance, measurements of
the colour factors and/or of $\alpha_s$ based on 4-jet
events\cite{colourfALEPH,fourjetsDELPHI,colourfOPAL}, studies of
the energy evolution of event shape distributions and of inclusive
charged particle production including measurements at the highest
energies compared to the prediction of hadronisation
models\cite{eventshapesALEPH,eventshapesDELPHI,eventshapesDELPHI2,eventshapesL3,ffOPAL},
measurements of the $b$ quark mass at the Z mass
scale\cite{bmassOPAL}. As the outcome of the work at LEP one can
conclude that the understanding of QCD phenomenology has much
improved and even rather subtle measurements are all consistent
with QCD predictions.

\section{Conclusion and Reflection}

It is appropriate now to recall what was known in summer 1989,
when LEP started and what was expected from LEP for the future.
Some examples of what was known are given below:

\[ \begin{array}{l}
m_Z = 91.12 \pm 0.16 \: GeV, \\ m_W = 80.0 \pm 0.36 \: GeV,
\\ sin^2\theta_W = 0.227 \pm 0.006, \\ N_{\nu} = 3.0 \pm
0.9.
\end{array} \]

It was expected, of course, that LEP would improve the accuracy
substantially. Looking back at the review talks presented by G.
Altarelli\cite{altarelli} at the Lepton Photon Symposium 1989 in
Stanford and by R. Barbieri\cite{barbieri} at the EPS Conference
1989 in Madrid one finds the expected experimental errors compared
in Table \ref{tab:achieved} with those actually achieved. I should
remark that the error for $N_{\nu}$ quoted as {\it expected} is
from the answer which was given by the DELPHI Collaboration to the
LEPC in 1982. In the end, all measurements turned out to be much
more precise than expected. Despite this precision the SM
continues to be in good shape.

Why was LEP so successful? Many fortunate facts had to come
together:\\ - A highly dedicated machine group responsible for the
excellent performance of LEP, \\ - low background in the
detectors,\\ - good performance of all detectors from the pilot
run in August 1989 till the end of data taking,
\\ - effective division of work between CERN and the outside
laboratories, \\- close cooperation between the 4 collaborations
and also between LEP and SLD (without avoiding competition),
\\- close cooperation between experiments and the machine group,
\\- and, very important, close cooperation with theory groups.

Many analyses are continuing and still more can be expected in the
future.

\begin{table}
\caption{Expected and achieved precision at
LEP.}\label{tab:achieved}
\begin{tabular}{|c|c|c|}

\hline

\raisebox{0pt}[12pt][6pt]{$Quantity$} &

\raisebox{0pt}[12pt][6pt]{$Expected \:\: error$} &

\raisebox{0pt}[12pt][6pt]{$Achieved$} \\

\hline

\raisebox{0pt}[12pt][6pt]{$m_Z$} &

\raisebox{0pt}[12pt][6pt]{50 to 20 $MeV$} &

\raisebox{0pt}[12pt][6pt]{\bf 2.1 $MeV$} \\

\raisebox{0pt}[12pt][6pt]{$m_W$} &

\raisebox{0pt}[12pt][6pt]{100 $MeV$} &

\raisebox{0pt}[12pt][6pt]{\bf 39 $MeV$} \\

\raisebox{0pt}[12pt][6pt]{$N_{\nu}$} &

\raisebox{0pt}[12pt][6pt]{0.3} &

\raisebox{0pt}[12pt][6pt]{\bf 0.008} \\

\raisebox{0pt}[12pt][6pt]{$A^{0,\mu}_{FB}$} &

\raisebox{0pt}[12pt][6pt]{0.0035} &

\raisebox{0pt}[12pt][6pt]{\bf 0.0013} \\

\raisebox{0pt}[12pt][6pt]{$A^{0,b}_{FB}$} &

\raisebox{0pt}[12pt][6pt]{0.0050} &

\raisebox{0pt}[12pt][6pt]{\bf 0.0017} \\

\raisebox{0pt}[12pt][6pt]{\(\mathcal A_{\tau}\)} &

\raisebox{0pt}[12pt][6pt]{0.0110} &

\raisebox{0pt}[12pt][6pt]{\bf 0.0043} \\

\hline
\end{tabular}
\end{table}

\section*{Acknowledgments}
First I would like to thank the organizing committee for giving me
the opportunity to present this summary talk. Preparing such a
talk is not possible without numerous communications with
colleagues from the LEP Collaborations. In particular I profited
much from discussions and mail exchanges with P. Antilogus, E.
Barberio, R. Chierici, M. Elsing, P. Gagnon, F. Glege, M.
Gr\"unewald, J. Holt, R. Jones, M. Kienzle, N. Kjaer, W. Liebig,
K. M\"onig, S. Myers, C. Parkes, A. Stocchi, H. Voss, Ch. Weiser
and D. Wicke. I am very grateful to G. Myatt for many helpful
suggestions and also for carefully reading the manuscript. I thank
S. Braccini for his assistance during the conference and during
finalising the manuscript.

\section*{Discussion}

{\it Alberto Sirlin, New York University}: I have an observation
and a question. \\ i) With respect to the evidence for genuine
electroweak corrections, I think there is a very simple argument
that shows a very large signal. It consists of measuring the
radiative correction $\Delta r$ by using the experimental results
for $m_W$ and $m_Z$, and comparing with the value $\Delta r$ would
have if the only contribution arose from the running of $\alpha$.
Last time I did this, about a year ago, I found a difference
amounting to many standard deviations. \\ ii) The question is:
what is the $\chi^2$ per degrees of freedom of the most recent
electroweak global fit? \\ \\ {\it J. Drees}: The most recent MSM
fit to all electroweak data including the direct measurements of
$m_W$ and $m_t$ has $\chi^2/ndf = 22.9/15$ corresponding to the
still reasonable probability of 8.6\%.

\end{document}